\documentclass[epjST]{svjour}
\usepackage{graphicx}
\usepackage{latexsym}
\usepackage{amsmath}
\usepackage{amssymb}
\usepackage{color}
\newcommand{\ket}[1]{| #1 \rangle}
\newcommand{\avg}[1]{\langle #1 \rangle}

\newcommand{\beq}{\begin{eqnarray}}
\newcommand{\eeq}{\end{eqnarray}} 

\newcommand{\Eq}[1]{\textcolor{blue}{Eq.\!\!~(\ref{#1})}}

\newcommand{\wj}{w_J}
\newcommand{\ei}{\hat{a}}
\newcommand{\eid}{\hat{a}^{\dag}}
\newcommand{\hn}{\hat{n}}

\newcommand{\Jx}{\hat{J}_x}
\newcommand{\Jy}{\hat{J}_y}
\newcommand{\Jz}{\hat{J}_z}

\begin{document}


\title{Suppression of collision-induced dephasing by periodic, erratic, or noisy driving}

\author{Christine Khripkov \inst{1} \and Amichay Vardi \inst{1} \and Doron Cohen \inst{2}}

\institute{Department of Chemistry, Ben-Gurion University of the Negev, Beer-Sheva 84105, Israel \and Department of Physics, Ben-Gurion University of the Negev, Beer-Sheva 84105, Israel}

\abstract{
We compare different driving scenarios for controlling the loss of single particle coherence
of an initially coherent preparation in the vicinity of the hyperbolic instability of the two-mode bose-Hubbard model. In particular we contrast the quantum Zeno suppression of decoherence by broad-band erratic or noisy driving,  with the Kapitza effect obtained for high frequency periodic monochromatic driving. 
} 

\maketitle

\section{Introduction}
\label{sec:intro}

The physics of Bose-Einstein condensates (BECs) confined in periodic lattice potentials
is captured in the tight binding approximation by the Bose-Hubbard Hamiltonian (BHH). 
With only two coupled condensates this model reduces 
to the Bose-Josephson form \cite{BHH1,BHH2,BHH3,BHH4,VardiAnglin1,VardiAnglin2}, aka the Bose-Hubbard dimer,
written in spin form as  
\beq \label{eq:BHH}
H \ \ = \ \ -\left[K-f(t)\right]\Jx+U\Jz^2~,
\eeq
where $K$ is the inter-mode coupling, which we later modulate by adding time dependent driving $f(t)$, and $U$ is the intra-mode interaction parameter.  
The spin operators are defined in terms of the bosonic 
creation and annihilation operators $\eid_i$ and $\ei_i$
of particles in mode ${i=1,2}$. 
Namely, $\Jz=(\hn_1-\hn_2)/2$ corresponds to the 
occupation difference, 
while $\Jx=(\eid_1\ei_2+\eid_2\ei_1)/2$ and $\Jy=(\eid_1\ei_2-\eid_2\ei_1)/(2i)$ correspond to the real and imaginary parts of the inter-mode coherence. 
Since the total particle number is conserved at $\hn_1+\hn_2=N$, the total spin is set to $j=N/2$. 
For simplicity we assume repulsive interaction $U>0$, noting that results for an attractive interaction are easily deduced by taking $K\rightarrow -K$ and $E\rightarrow-E$. 

The characteristic interaction parameter of the Hamiltonian (\ref{eq:BHH}) is
\beq
u \ \ \equiv \ \ NU/K~.
\eeq
Single particle coherence, the hallmark of a global macroscopic order, 
is determined by the length of the Bloch vector $\vec{S}=\avg{\vec{J}}/j$. 
The classical limit, which preserves ${S=1}$,  
is obtained by restricting the quantum Hilbert space 
to the subset of {\em spin coherent states},
\begin{equation}
\label{SCS}
|\theta,\varphi\rangle \ \ \equiv \ \ \exp({-i\varphi\Jz})\exp({-i\theta\Jy}) \ |j,j\rangle~,
\end{equation}
where $n=\cos \theta$ is the relative population imbalance, 
and $\varphi$ the relative phase between the two condensates. 
To the extent that an initial spin coherent state evolves 
only to other coherent states, the dynamics can be described 
by the mean-field equations, where the spin operators 
in Eq.~(\ref{eq:BHH}) are replaced by $c$-numbers with ${\cal O}(1/N)$ accuracy.

\begin{figure}
\centering
\includegraphics[width=\textwidth]{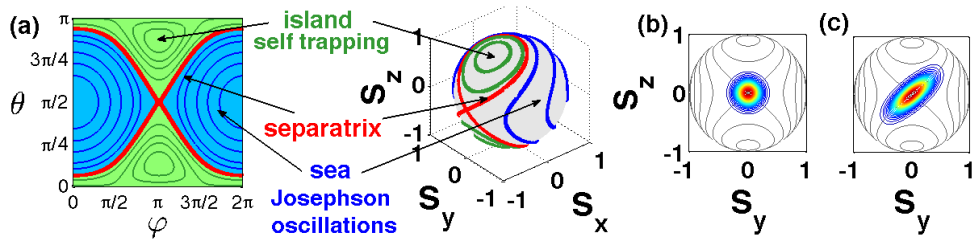}
\caption{(a) Phase-space structure of the BHH for $u=2.5$. 
Lines depict equal energy contours. 
A separatrix with an isolated hyperbolic point 
at $\vec{S_\pi}=\ket{\pi/2,\pi}$ separates the 
``islands" and ``sea" regions. 
The Wigner distribution for the $\vec{S_\pi}$ 
coherent state is shown in (b), and for a squeezed state in (c).}
\label{fig1}
\end{figure}  

The equal energy contours of the classical dimer Hamiltonian in the Josephson interaction regime $1<u<N^2$ are illustrated in Fig.~\ref{fig1}. As seen, the spherical phase-space consists of three distinct regions of motion: Two nonlinear ``islands" are separated from a ``sea region" 
by a separatrix. The sea trajectories correspond to Rabi-Josephson oscillations of the population \cite{BHH1,Oscillations1,Oscillations2}, whereas the island trajectories correspond to self-trapped motion \cite{BHH2,SelfTrap1,SelfTrap2}.

Linearization of the dimer BHH around the {\em elliptic} fixed-point $\vec{S_0}=(1,0,0)$, corresponding to the spin coherent state $|\pi/2,0\rangle$, 
gives the Josephson frequency $\mbox{$\omega_J=K\sqrt{u+1}$}$ 
of small oscillations around the ground state. 
By contrast, linearization around the {\em hyperbolic} fixed-point $\vec{S_\pi}=(-1,0,0)$, corresponding to the spin coherent state $|\pi/2,\pi\rangle$, 
gives the squeezing rate \cite{Khodorkovsky2},
\beq
\wj \ \  = \ \ K\sqrt{u-1}~.
\eeq
Note that in the Josephson regime $u>1$ implies the appearance of the separatrix and hence the hyperbolic behavior of  $\vec{S_\pi}$.  We also assume that $u\ll N^2$, so that the ``sea region" and the hyperbolic structure are well resolved.
 
The phase-space Wigner distribution of an initial coherent preparation $|\pi/2,\pi\rangle$ is approximately a minimal Gaussian centered at $\vec{S_\pi}$, with uncertainty radius $\mbox{$r_0=\sqrt{2/N}$}$ (Fig.~\ref{fig1}b). Due to the hyperbolic dynamics in the absence of driving, this Gaussian undergoes expansion and contraction with factors $e^{\pm \wj t}$, respectively, along two principal axes (Fig.~\ref{fig1}c) \cite{Khodorkovsky2}.  The angle between the principal axes is twice the value of $\Theta=\tan^{-1}(\wj/K)$.

Due to the growth of variance along the expanding axis of the squeezed Gaussian state, the one-particle coherence $S$ decreases. Using a simple phase space picture, a good approximation for the loss of coherence is given by \cite{Khripkov},
\beq \label{eq:squeezing}
S(t) \ \ = \ \ \exp\left\{-\frac{1}{2}\left(\avg{r^2}-r_0^2\right) \right\}
\ \ = \ \ \exp\Big\{ -r_0^2 \cot^2(2\Theta)\sinh^2(\wj t) \Big\}~.
\eeq
where $\avg{r^2}$ is the angular spreading of the squeezed Gaussian. One implication of this squeezing process is the hyperbolic growth of deviations from mean-field theory \cite{VardiAnglin1,VardiAnglin2,BoukobzaChuchem}. Equation~(\ref{eq:squeezing}) establishes a short-time one-to-one connection between the fluctuations ratio along the principal axes $e^{2 \wj t}$ and the resulting one-particle coherence $S$. Due to the finite phase-space volume, the short time squeezing is then followed by a series of revivals and collapses due to the repeated
folding and interference of the Wigner distribution \cite{Revivals1,Revivals2,Revivals3,Revivals4,Chuchem10,BoukobzaChuchem}. Of course, at these later times the phase-space Wigner distributions are no longer squeezed Gaussians and $S$ no longer reflects the degree of squeezing.

\section{Driving scenarios}
\label{sec:driving}

In this work we compare three strategies for the stabilization of $\vec{S_\pi}$ coherent preparation by introducing a time-dependent driving $f(t)$ as in \Eq{eq:BHH}, i.e. by the modulation 
of the two-mode coupling strength. The simplest driving scheme
assume a periodic mono-chromatic sinusoidal time dependence,   
\beq
\label{eq:harmonic}
f(t)=D\sin(\Omega t+\phi)~,  \ \ \ \ \ \ \ \mbox{[harmonic driving]}
\eeq
with intensity $D$ and frequency $\Omega$. 
Defining dimensionless frequency $\tilde{\Omega}\equiv \Omega/\omega_J$, 
and dimensionless driving strength $q\equiv\sqrt{u}D/\Omega$,  
fast and slow driving correspond to $\tilde{\Omega}\gg1$ and $\tilde{\Omega}\ll1$,
respectively, whereas $q\gg\tilde{\Omega}$ and $q\ll\tilde{\Omega}$ 
correspond to strong and weak driving.
 
Next we consider a broad-band erratic driving field $f(t)$, with zero average and a short correlation time. This can be viewed  as a {\em realization} of a stochastic process, such that upon averaging
\beq
\label{e7}
\avg{f(t)f(t')}=2D\delta(t-t')~,
\ \ \ \ \ \ \ \mbox{[erratic driving]}
\eeq
We stress that `erratic driving' still refers to {\em Hamiltonian} evolution with a deterministic, well controlled  $f(t)$. An experimentalist can reproduce the same $f(t)$ many times, and perform a quantum measurement of the outcome. Optionally, he can carry out experiments with different realizations of $f(t)$, and accumulate statistics. 

Finally, we investigate the case of full-fledged {\em quantum noise}, induced by the coupling of the system to a ``bath". In this case averaging is not at the courtesy of the experimentalist, but an essential ingredient in the proper description of the reduced dynamics. Assuming that the quantum noise has the same correlation as postulated in \Eq{e7}, the probability matrix would obey 
the master equation,
\beq \label{eq:master}
\dot{\rho}=-i[H,\rho]-D[\Jx,[\Jx,\rho]]~,
\ \ \ \ \ \ \ \mbox{[noisy driving]}
\eeq
where the bath degrees of freedom have been traced out. This is equivalent to averaging the Hamiltonian erratic driving dynamics over {\em all} possible realizations of $f(t)$.

\section{Harmonic driving - Kapitza effect}
\label{sec:kapitza}

In order to understand the effect of harmonic driving, we note that for small population imbalance $n\ll N$ (i.e. when $\theta\approx\pi/2$) 
we have $J_x\approx(N/2)\cos\varphi$, 
and the BHH takes the form of a pendulum Hamiltonian, 
\beq
H \ \ = \ \ Un^2- (1/4)NK\cos{\varphi}~.
\eeq
The stable ${\bf S_0}$ and the unstable ${\bf S_\pi}$ fixed points corresponds to the `down' and `up'  directions of the pendulum. Within this approximation, off-resonant high-frequency weak drive, corresponding to vertical driving of the pendulum axis, results in an effective potential ${V^{\text{eff}}=(1/4)q^2 Kj\sin^2(\varphi)}$ \cite{Kapitza1,Kapitza2,Kapitza3,Kapitza4}. Sufficiently strong driving (${q^2 > 2}$)  stabilizes the $\varphi=\pi$ fixed-point, producing the Kapitza inverted pendulum effect.

\begin{figure}
\centering \includegraphics[width=\textwidth] {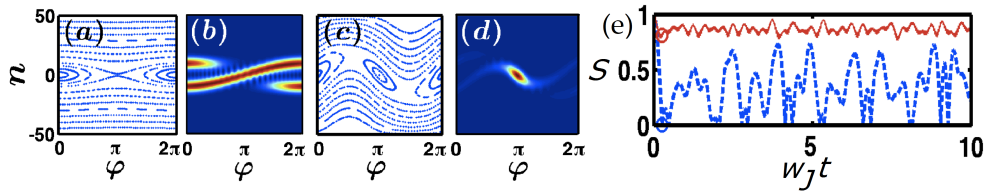}
\caption{Quantum Kapitza dynamics: (a) classical trajectories without driving ($q=0$); 
(b) corresponding Husimi distribution after drive-free BHH propagation for $\wj t \approx 0.25$, assuming $\vec{S_\pi}$ coherent preparation; (c) stroboscopic classical trajectories with high frequency harmonic driving ($q=3,\tilde{\Omega}=30$); (d) corresponding Husimi distribution after driven BHH propagation of the same preparation for the same time as in (b).  Panel (e) shows the corresponding
one-particle coherence dynamics with (solid red line) and without (dashed blue line) driving. Circles mark the time at which the Husimi distributions
(b,d) are plotted. Parameters are $u=100$, $N=100$. }
\label{fig2}
\end{figure}

This analysis was extended to the full spherical phase space and general driving term \cite{Boukobza}. Given that the Hamiltonian is of the form ${H +f(t)W}$, the effective potential is $-[1/(4\Omega^2)] \ [W,[W,H]]$, which for $W=\Jx$ contributes a term proportional to $\sin^2\theta\sin^2\varphi$, as well as additional terms that slightly renormalize the bare values of $K$ and $U$.  This effective potential leads to the stabilization of the $\vec{S_\pi}$ point, and correspondingly, to the suppression of collisional dephasing as shown in Fig.~\ref{fig2}.

\section{Erratic driving - Squeezing axis randomization}
\label{sec:erratic}

Fig.~\ref{fig3} shows the dynamics of the single particle coherence 
under the influence of erratic driving. In contrast to the drive-free 
dynamics (\ref{eq:squeezing}), the orientation of the wavepacket
relative to the principal squeezing axes is not constant: the erratic driving 
randomizes this orientation on time scale $t_D=1/(2D)$, 
resulting in angular diffusion \cite{Khripkov}. The squeezing rate 
is no longer constant, thus Eq.~(\ref{eq:squeezing}) should be 
replaced by the general expression  
\beq \label{eq:lambda}
S(t)=\exp\{-r_0^2 \sinh^2(\Lambda)\}~,
\eeq
with a squeezing parameter $\Lambda$ that accumulates 
in a stochastic fashion. The accumulated squeezing 
parameter $\Lambda$ in a given realization of $f(t)$ is a sum of uncorrelated 
variables, and therefore, according to the central limit theorem, 
its many-realization distribution is normal, with some mean value $\mu$ 
and dispersion $\sigma$. It follows from \Eq{eq:lambda}
that the corresponding distribution of $S$ is log-wide, with typical value 
(median) that might be very different from the mean:
\begin{align} \label{eq:median}
S_{\text{median}} &= \exp\Big\{-r_0^2 \sinh^2(\mu) \Big\}~,
\\
\label{eq:average}
S_{\text{mean}} &\approx  
\exp\left\{ -\frac{r_0^2}{2}\left[e^{2\sigma^2}\cosh(2\mu)-1 \right]\right\}~.
\end{align}

\section{Noisy driving - Quantum Zeno effect}
\label{sec:noise}

Since quantum noise may be viewed as an average over all realizations of erratic driving, the expected decay in this case should follow \Eq{eq:average}. 
We would like to obtain a more practical version of this 
formula using a slightly different procedure 
that is inspired by the analysis of 
the Quantum Zeno effect \cite{QZE1,QZE2,QZE3,QZE4,QZE5}.
Previous analysis \cite{Khodorkovsky1,Khodorkovsky2} 
led to Fermi-Golden-rule type expression 
for the decay of coherence:
\beq
S=\exp[-r_0^24D_w t]~, \ \ \ D_w=[\cot^2(2\Theta)]\frac{w_J^2}{8D}~,
\label{QZEL}
\eeq
which is valid for ${t_D\ll (1/w_J)}$ and restricted to short times such that $D_w t \ll1$.

\begin{figure}
\centering
\includegraphics[width=\textwidth]{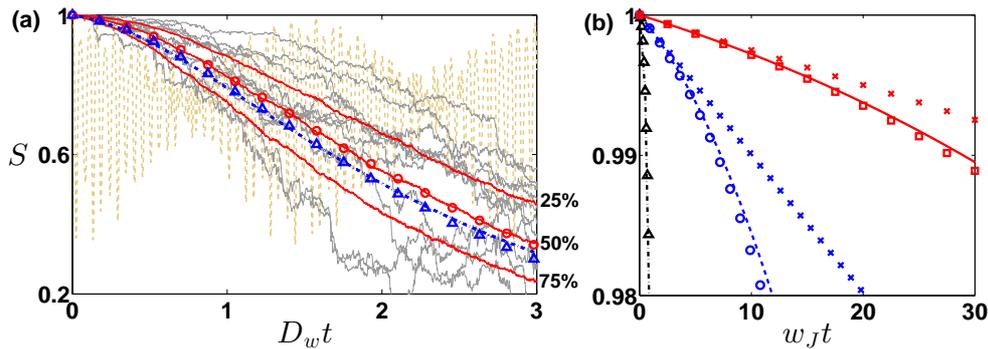}

\caption{
The one-body coherence $S$ is plotted as a function of time, 
for $N=50$ particles and $u =2$.
On the left: the dashed/yellow line is the drive-free decay. 
Single realizations of erratic driving 
whose intensity is $D=37.5\wj$ are illustrated by gray lines. 
The red lines mark the 25\%, the 50\% (median), 
and the 75\% percentiles over 2000 realizations. 
The dash-dotted/blue line is the average 
as obtained from the master equation Eq.~(\ref{eq:master}).
The symbols are based on Eq.~\ref{eq:median} ($\circ$)   
and on Eq.~\ref{eq:average} ($\triangle$).
On the right: the theoretical predictions are tested 
for shorter times during which coherence is not yet lost.
The parameters are $N=100$ and $u=2$.
The noise-free decay is marked by the dash-dotted/black line, 
while the dashed/blue is for $D=10\wj$, 
and the solid/red is for $D=40\wj$.
The symbols are based on Eq.~\ref{eq:squeezing} ($\triangle$), Eq.~\ref{QZEL} ($\times$), and Eq.~\ref{eq:qze} ($\circ$, $\square$).
}
\label{fig3}
\end{figure}

Our analysis does not assume that $D_w t \ll1$, 
but we still assume that $\mbox{$D_w t \ll \log(N)$}$, 
such that $S$ is not far from unity 
and the folding of the wavepacket around the 
sphere can be ignored.
During each squeezing interval $t_D$ 
the angular variance grows as 
$\avg{r^2}_{t+t_D}=\avg{r^2}_t [1+2\cot^2(2\Theta)\sinh^2(\wj t_D)]$, 
before being reset by the noise. 
Approximating the $\sinh^2$ by a quadratic function, 
we get after $t/t_D$ steps an exponentially 
growing $\avg{r^2}_t$, leading to 
\beq \label{eq:qze}
S(t)=\exp\left\{ -\frac{r_0^2}{2} [\exp(8D_w t)-1]\right\}~.
\eeq
The accuracy of this expression 
is demonstrated in the right panel of Fig.~\ref{fig3}.

\section{Summary}
\label{sec:summary}

We have studied the effect of driving on the hyperbolic instability of the Bose-Hubbard dimer. High frequency off-resonant harmonic driving results in the many-body equivalent of the Kapitza pendulum effect. On the other hand, erratic driving and quantum noise slow down 
the loss of single particle coherence via a many-body quantum Zeno effect. 
However, for long times the decay of $S$ departs 
from the Fermi-golden-rule expectation.
Namely, the interplay of angular diffusion with the  
hyperbolic squeezing results in log-wide statistics for $S$,  
whose typical value differs from the algebraic mean.

\begin{acknowledgement}
This research was supported by the Israel Science Foundation (Grants No.346/11 and No.29/11) 
and by the United States-Israel Binational Science Foundation (BSF).
\end{acknowledgement}


\end{document}